\documentclass[11pt,a4paper]{article}
\usepackage[makeroom]{cancel}
\usepackage{bm}
\usepackage{amsmath}
\usepackage{amsfonts}
\usepackage{amssymb}
\usepackage{graphicx}
\usepackage{authblk}
\usepackage{cite}
\usepackage[left=2.4cm,top=2.4cm,right=2.4cm,bottom=2.4cm,nohead]{geometry}
\DeclareGraphicsExtensions{.png,.jpg,.pdf}
\usepackage{epstopdf}
\usepackage{float}
\usepackage{fouriernc} 
\usepackage{enumitem}
\usepackage[utf8]{inputenc}
\usepackage[english]{babel}
\usepackage{mathtools}
\usepackage{amscd}
\usepackage{mathrsfs}
\usepackage{amsthm}
\usepackage{siunitx}
\usepackage{caption}
\usepackage{subcaption}
\setlist{nolistsep,leftmargin=*}

\usepackage{mhchem} 
\usepackage{cleveref} 
\usepackage{xcolor}
\usepackage{soul}
\usepackage{mdframed}
\usepackage{framed}
\usepackage{nicefrac}
\usepackage{longtable}
\DeclareMathAlphabet{\mathpzc}{OT1}{pzc}{m}{it}

\title{Extended Donnan model for ion partitioning \\ in charged nanopores} 

\makeatletter

\renewcommand\AB@authnote[1]{\textsuperscript{\normalfont#1}}

\makeatother

\author[1]{R. Wang,}
\author[1]{M. Elimelech,}
\author[2]{P.M. Biesheuvel} 

\affil[1]{Department of Chemical and Environmental Engineering, Yale University, 
United States of America.}
\affil[2]{Wetsus, European Centre of Excellence for Sustainable Water Technology, Leeuwarden, The~Netherlands.}

\date{} 

\newcommand{\s}[1]{\mathrm{_{#1}}}

\begin{document}

\maketitle

\begin{abstract}

Membranes consist of pores and the walls of these pores are often charged. In contact with an aqueous solution, the pores fill with water and ions migrate from solution into the pores until chemical equilibrium is reached. The distribution of ions between outside and pore solution is governed by a balance of chemical potential, and the resulting model is called a Donnan theory, or Donnan equation. Including a partitioning coefficient that does not depend on salt concentration results in an extended Donnan equation `of the first kind'. Recently, an electrostatic model was proposed for ions in a pore based on the arrangement of ions around strands of polymer charge, including also ion activity coefficients in solution. That framework leads to an extended Donnan equation `of the second kind', which has extra factors depending on ion concentrations in the pores and salt concentration in solution. In the present work, we set up another Donnan model of the second kind by evaluating the Coulombic interactions of ions in a cylindrical pore, including the interaction of ions with the charged pore walls and between the ions. We assume that counterions are near the pore wall while coions distribute over the center region. Starting from a complete analysis, we arrive at an elegant expression for the chemical potential of ions in such a pore. This expression depends on coion concentration, pore size, and other geometrical factors, but there is no additional dependence on counterion concentration and charge density. This model predicts the Coulombic contribution to the chemical potential in the pore to be small, much smaller than predicted by the electrostatic model from literature. Instead, we predict that up to around 1 M salt concentration, activity effects of ions in solution are more important.  

 
\end{abstract} 

\section{Introduction}

When a charged porous material, such as a membrane, is in contact with an aqueous solution, water fills the pores of the material. Molecules, charged and uncharged, distribute between the external solution and the pores of the material. For a salt solution and charged membrane pores, the Donnan equation is the classical theory to describe ion partitioning, and the related Donnan potential~\cite{Schlogl_1966,Sonin_1976, Galama_2013,Kamcev_2015,Kamcev_2016,Biesheuvel_Dykstra_2020,Kitto_2022,Biesheuvel_2022,Biesheuvel_2023}. Donnan theory is of importance to describe charged porous materials such as ion absorbers, hydrogels, and membranes for water desalination. Besides the electrostatic effect related to maintaining electroneutrality in the material, also important is an additional contribution described by a partitioning coefficient $\Phi_i$, that can be due to an affinity of ions to the membrane, or for instance volume exclusion. This term is included in the extended Boltzmann equation (an equation which applies to individual ions), but is generally left out in an analysis of the Donnan balance. This contribution, however, must also be included in a Donnan balance.

The partitioning coefficient, $\Phi_i$, depends on the type of ion, the membrane material, and structural properties such as porosity and pore size, but it does not depend on ion concentrations inside or outside the membrane. Instead, it simply operates as an additional constant in the Donnan balance. We call this type of theory an extended Donnan model `of the first kind', and it works very well to describe ion absorption in porous media, also for salt mixtures. To describe the constant factor $\Phi_i$, many physical-chemical forces between ions, solvent, and membrane must be considered that all may play a role~\cite{Wright_2007}. Theories developed for membrane applications, such as ion dehydration and dielectric exclusion, are helpful to find a proper value for $\Phi_i$.  

\begin{figure}
\centering
\includegraphics[width=0.85\textwidth]{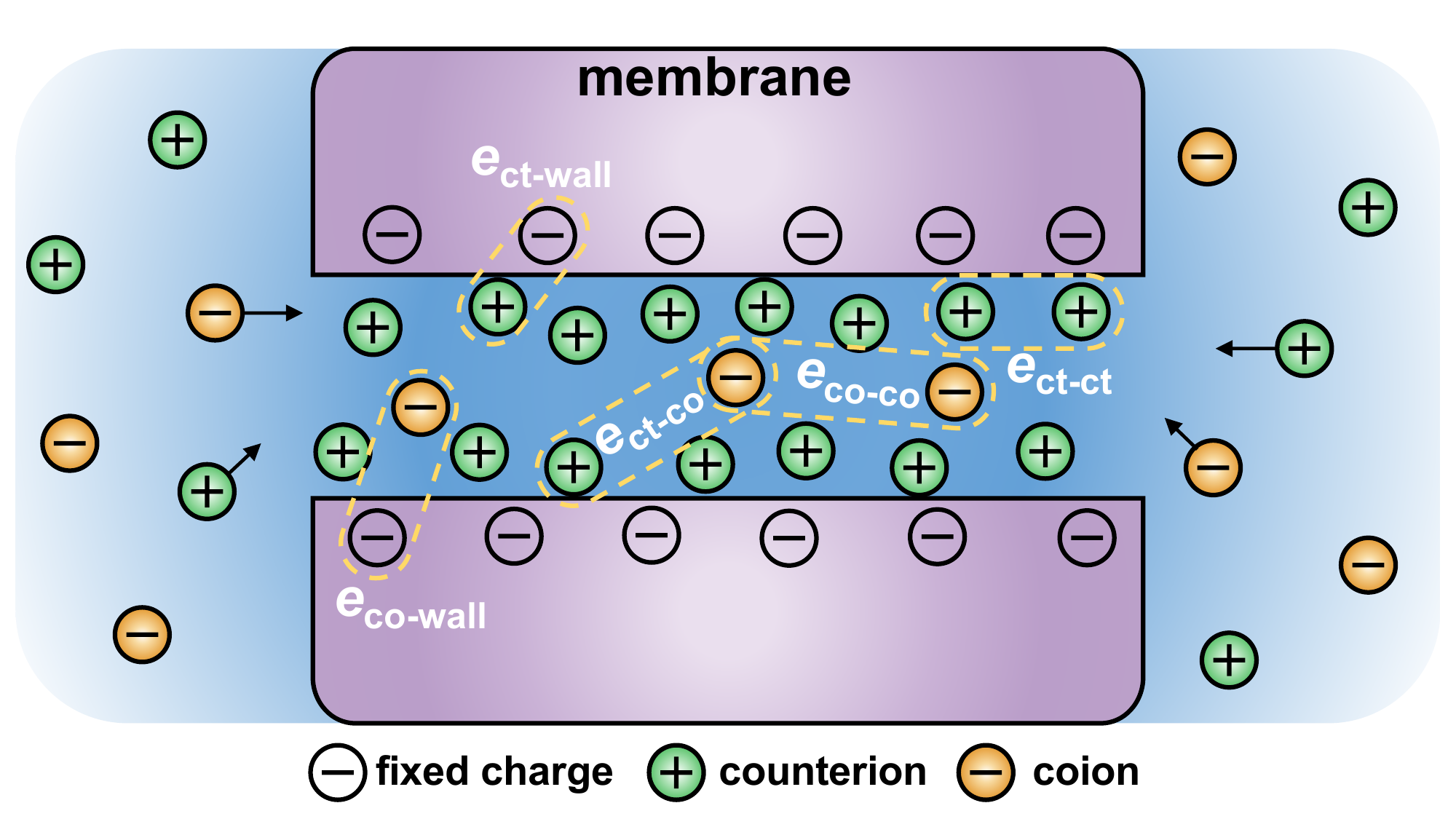}
\vspace{-0pt}
\caption{Schematic of ion distribution in a charged nanopore inside a porous membrane. With fixed charges on the pore walls (here of negative charge sign), we have counterions residing near the pore wall and coions in the center region of the pore. To account for ion activity in the pore, all five mutual interaction energies between ions and pore charge are considered.}
\label{fig_scheme}
\end{figure}

About 10 years ago, a new theory was proposed based on the theory of ion condensation around polymer chains, which introduces an additional dependence of the Donnan equation on ion concentrations in the membrane, and also implements the activity coefficient of ions in the external solution~\cite{Kamcev_2015,Kamcev_2016,Kitto_2022}. We call this an extension of the Donnan equation `of the second kind', because the modification to the original Donnan model is now dependent on ion concentrations, both inside and outside the membrane. However, evaluating this model it is found that the dependence of this additional effect on salt concentration is not very strong, and thus in practice this theory is not very different from an extended Donnan model of the first kind (those are models based on a fixed $\Phi_i$). 
The ion condensation theory is based on a geometry where ions are distributed around charged polymer chains, but that may not be the best representation of how ions arrange in a charged nanopore. In this theory, the condensation effect is strong, and stronger than the activity effect of ions in solution, resulting in a significant attraction of ions into the pores of a charged material.

In the present work, we discuss a different theory of the Coulombic forces between ions in a charged cylindrical nanopore. We evaluate this theory for the influence of external salt concentration, ion and pore sizes, and membrane charge on ion partitioning. This analysis contributes to resolving the question whether an extended Donnan theory of the first kind is sufficient, or if a theory of the second kind is necessary. In this paper, we focus on a 1:1~salt solution, and we explicitly consider all Coulombic forces between the ions in the pore, and their interaction with the charged pore wall, see Fig.~\ref{fig_scheme}. Coulombic models have proven to be very effective to describe ion activity coefficients in solution~\cite{Biesheuvel_2020}, which suggests that a similar approach can also be applicable for the distribution of ions in a pore. 
Because we make use of Coulomb's law between all ions, all energies are additive, and expressions can be developed for the mean activity coefficient of an ion in the membrane. We implement this equation in a Donnan balance and evaluate whether a significant effect of salt concentration on the total partitioning is predicted (which is then an extended Donnan model of the second kind), or whether we arrive at a Donnan model of the first kind, where a constant factor $\Phi_i$ suffices.

\section{Theory of ion activity coefficients in a charged nanopore}

In the present section we derive an expression for the mean activity coefficient of ions in a charged nanopore, $\gamma_{\text{m},\pm}$, for given values of the counterion and coion concentration inside the pore, $c\s{ct}$ and $c\s{co}$. We consider a 1:1 salt solution with both ions able to enter the pores of the membrane. In the pore we assume that counterions reside near the pore wall (to which they are strongly attracted), all with their center at, or near, a radial coordinate $r\s{ct}$ (less than the pore radius $r\s{p}$), while coions distribute homogeneously in a center region between the pore center axis and a coordinate $r\s{co}$ (less than $r\s{ct}$). In a later section this expression is combined with an overall model that also includes local electroneutrality, ion activity in the outside solution, and an additional constant partitioning coefficient, $\Phi_i$, to evaluate the Donnan equation for the distribution of ions between a solution and the pores of a charged material.

We assume we have a pore of radius $r\s{p}$ (and thus cross-sectional area $A = \pi r_\text{p}^2$) with counterions at concentration $c\s{ct}$ (all concentrations are defined per unit pore volume) and coions at $c\s{co}$. The charge density per unit pore volume is \textit{X} with unit mM, and per unit pore area it is $\sigma$ with unit~m\textsuperscript{-2}. These definitions for membrane charge are related by
\begin{equation}
 2 \sigma  =  X \,  N\s{av} \, r\s{p} 
 \label{eq_sigma_X}
\end{equation}
where $N\s{av}$ is Avogadro's number. Charge neutrality in the membrane pores is described by~\cite{Schlogl_1966}
\begin{equation}
z\s{ct} c\s{ct} + z\s{co} c\s{co} +  X = 0 
\label{eq_EN}
\end{equation}
with $z_i$ the valency of co- and counterions. Note that in the equations below, charge $\sigma$ is assumed to be a positive quantity. 

To evaluate the Coulombic contribution to the chemical potential of an ion in a charged nanopore, and consequently to calculate the activity coefficient, we have five energies to consider: coion-wall,  coion-counterion, coion-coion, counterion-wall, and counterion-counterion. Each energy ${e}_i$ can be recalculated to a contribution to the chemical potential, $\mu_i$. Below, these five contributions are labelled ${\mu}_1$ to ${\mu}_5$. Three out of the five are repulsive, i.e., will be a positive contribution, and two are attractive, which are those between counterions and coions, and between counterions and the pore wall.

In this calculation we derive expressions for the energy per ion, ${e}$, which can be recalculated to a contribution to the chemical potential, $\mu_i$, and each of these chemical potentials is equivalent to a contribution to the natural logarithm of the mean activity coefficient of ions in the membrane, $\ln \gamma_{\text{m},\pm}$. We note that it is not energies \textit{e} that directly translate to a contribution to $\ln\gamma_\pm$. These energies and potentials, \textit{e} and $\mu_i$, are dimensionless, i.e., can be multiplied by \textit{RT} to obtain a dimensional energy (potential) with unit J/mol.  

The dimensionless Coulombic energy between two unitary charges (brought together to a distance \textit{x} from infinitely far away) is ${e}=  \lambda\s{B}/ x $, where $\lambda\s{B}$ is the Bjerrum length (at room temperature, $\lambda\s{B} \sim 0.7$~nm)~\cite{Wright_2007,Biesheuvel_2020}. For the interaction of one discrete (unitary) charge interacting with a differential surface area d\textit{A} that has charge $\sigma$ (in m\textsuperscript{-2}), this energy is ${e}= \lambda\s{B} \, \sigma \, \int_S   x^{-1}  \; \text{d}A $. 

\subsection{coion-wall interaction}

The first of the five terms is the interaction of coions with the pore wall. 
For this term, the interaction energy with the pore wall is
\begin{equation}
e\s{co-wall} =  {4 \lambda\s{B}r\s{p}}  \sigma  {r_\text{co}^{-2}}   \, \int_0^{r\s{co}} \int_{-\pi}^{\pi} \int_{0}^\ell   {r}  \left( r_\text{p}^2-2 r\s{p} r \cos{\vartheta}+r^2+z^2 \right)^{-1/2} \; \text{d} z  \text{d} \vartheta   \text{d}r
\label{eq_e_co_wall}
\end{equation}
where $\ell$ is a distance in axial direction (counting from coordinate $z \! = \! 0$), up to where we evaluate the interaction. In Eq.~\eqref{eq_e_co_wall} one prefactor 2 is because we must integrate over interactions to the left and right of the key ion that is located at $z \! = \! 0$. 

Integration with \textit{z} leads to
\begin{equation}
e\s{co-wall} = {8 \lambda\s{B}r\s{p}}  \sigma  {r_\text{co}^{-2}}  \, \int_0^{r\s{co}} \int_{0}^{\pi}  {r} \; \ln \left(\frac{\ell + \sqrt{r_\text{p}^2-2 r\s{p}r\cos{\vartheta}+r^2+\ell^2}}{\sqrt{r_\text{p}^2-2 r\s{p}r\cos{\vartheta}+r^2}}  \right) \;   \text{d} \vartheta  \text{d}r \, .
\label{eq_e_co_wall_2}
\end{equation}
For large $\ell$, Eq.~\eqref{eq_e_co_wall_2} simplifies to
\begin{equation}
e\s{co-wall} ={8 \lambda\s{B}r\s{p}}  \sigma {r_\text{co}^{-2}}    \,  \int_0^{r\s{co}}   {r} \; \left\{ \pi \ln \left( \frac{2 \ell}{r\s{p}} \right) - \tfrac{1}{2} \int_0^\pi \ln\left(\rho^2-2 \rho \cos\vartheta + 1 \right) \text{d}\vartheta \right\}  \text{d}r 
\label{eq_e_co_wall_3}
\end{equation}
where $\rho=r/r\s{p}$. The integral over $\vartheta$ is zero, and thus Eq.~\eqref{eq_e_co_wall_3} simplifies to
\begin{equation}
e\s{co-wall} = {4 \pi \lambda\s{B}r\s{p}} \sigma    \ln \left( \frac{2 \ell}{r\s{p}} \right)  \, .
\label{eq_e_co_wall_4}
\end{equation}
This result equals the energy for the situation that all coions are at the center axis, and it makes sense that these numbers are indeed the same. (The energy in case of all coions located on the centerline, we derived by a different route.) This equivalence is analogous to the case of gravitational interaction. Because in a gravity calculation, if two objects have the same center of mass (same position and same mass), the interaction with another body (that is external to it) is the same.

We obtain a contribution to the chemical potential of the ions in the pore, $\mu_i$, according to $\mu_i = \sum_k \partial f / \partial c_k $, where $f$ is the free energy density given by $f=e_ic_i$, and the summation over $k$ indicates we add up a contribution from differentiation with respect to the anion and cation, while in the differentiation the concentration of the other ion is fixed. Following this procedure, for the first contribution, $\mu_1$, we find that it becomes the same as $e\s{co-wall}$ given by Eq.~\eqref{eq_e_co_wall_4}. 

\subsection{coion-counterion interaction}

Next we analyse the interaction of coions with the counterions. When considered from the viewpoint of the coions, this is analogous to the coion-wall interaction discussed above. Instead of Eq.~\eqref{eq_e_co_wall_4} we now arrive at
\begin{equation}
e\s{co-ct} = - 2    c\s{ct}    \alpha     \ln \left( \frac{2 \ell}{r\s{ct}} \right) 
\label{eq_e_co_ct}
\end{equation}
which is slightly different from Eq.~\eqref{eq_e_co_wall_4} because we replaced $r\s{p}$ with $r\s{ct}$, $\sigma$ with $c\s{ct} r_\text{p}^2/\left(2 r_\text{ct}\right) N\s{av}$, see Eq.~\eqref{eq_sigma_X}, and added a minus sign because this interaction is attractive. We introduce the constant $\alpha$, given by $\alpha=\lambda\s{B} A N\s{av}$, where pore area $A$ is given by $A=\pi r_\text{p}^2$. We multiply Eq.~\eqref{eq_e_co_ct} with $c\s{co}$, and this free energy density, \textit{f}, is differentiated with respect to $c\s{ct}$ and $c\s{co}$, resulting in 
\begin{equation}
\mu_2 =  - 2\alpha \left(c\s{ct} + c\s{co} \right) \ln \left(\frac{2 \ell }{r\s{ct}}  \right) \, .
\label{eq_mu_co_ct}
\end{equation}

\subsection{coion-coion interaction}

The third interaction is that between coions. They are distributed in the center region, between $r \! = \! 0$ and $r \! = \! r\s{co}$, and to find the energy per coion we integrate
\begin{equation}
e\s{co-co} =  \frac{2 \alpha c\s{co} }{\pi r_\text{co}^4}    \, \int_0^{r\s{co}} \int_0^{r\s{co}} \int_{-\pi}^{\pi} \int_{0}^\ell   \frac{r_1 r_2}{\sqrt{ r_\text{1}^2+r_2^2-2 r_1 r_2 \cos{\vartheta}+z^2}} \; \text{d} z  \text{d}\vartheta   \text{d}r_1 \text{d}r_2 \, .
\label{eq_e_co_co}
\end{equation}
Integration with \textit{z} for large $\ell$ leads to
\begin{equation}
e\s{co-co} =  \frac{4 \alpha c\s{co} }{\pi r_\text{co}^4}    \, \int_0^{r\s{co}} \int_0^{r\s{co}} r_1 r_2 \int_{0}^{\pi} \ln \left(\frac{2\ell}{\sqrt{r_\text{1}^2-2 r\s{1}r\s{2}\cos{\vartheta}+r_\text{2}^2}}  \right)  \text{d}\vartheta   \text{d}r_1 \text{d}r_2 \, .
\label{eq_e_co_co_intl}
\end{equation}
We can split the integral over $r_1$ into two parts for the convenience of integrating over $\vartheta$
\begin{multline}
e\s{co-co} =  \frac{4 \alpha c\s{co} }{\pi r_\text{co}^4} \left\{ \vphantom{\frac{\int^{A^A}}{\int^{B^B}}} \int_0^{r\s{co}} \int_0^{r\s{2}} r_1 r_2 \left(  \pi \ln \left( \frac{2 \ell }{r\s{2} } \right) - \frac{1}{2} \int_0^\pi \ln \left( \rho_\text{12}^2-2 \rho\s{12} \cos{\vartheta}+1  \right) \text{d} \vartheta   \right)   \text{d}r_1 \text{d}r_2 \right. \\
\left.+   \int_0^{r\s{co}} \int_{r\s{2}}^{r\s{co}} r_1 r_2 \left(  \pi \ln \left( \frac{2 \ell }{r\s{1} } \right) - \frac{1}{2} \int_0^\pi \ln \left( \rho_\text{21}^2-2 \rho_\text{21} \cos{\vartheta}+1  \right) \text{d} \vartheta   \right)   \text{d}r_1 \text{d}r_2  \vphantom{\frac{\int^{A^A}}{\int^{B^B}}}  \right\}
\label{eq_e_co_co_part1}
\end{multline}
where $\rho\s{12}=r\s{1}/r\s{2}$ and $\rho\s{21}=r\s{2}/r\s{1}$. Noticing that the integrals over $\vartheta$ are zero, Eq.~\eqref{eq_e_co_co_part1} simplifies to
\begin{equation}
e\s{co-co} =  \frac{4 \alpha c\s{co}}{r_\text{co}^4}   \left\{  \vphantom{\frac{\int^{A^A}}{\int^{B^B}}} \int_0^{r\s{co}} r_2   \ln \left( \frac{2 \ell }{r\s{2} } \right) \int_0^{r\s{2}} r_1 \text{d}r_1 \text{d}r_2 + \int_0^{r\s{co}} r_2 \int_{r\s{2}}^{r\s{co}} r_1  \ln \left( \frac{2 \ell }{r\s{1} } \right)   \text{d}r_1 \text{d}r_2 \right\} 
\label{eq_e_co_co_part1-2}
\end{equation}
which simplifies further to
\begin{equation}
e\s{co-co} =  \alpha c\s{co} \left( \ln \left(\frac{2 \ell }{r\s{co}} \right) + \frac{1}{4} \right) \, .
\label{e_co-co_final}
\end{equation}
We multiply Eq.~\eqref{e_co-co_final} with $c\s{co}$ and take the derivative of this energy density, \textit{f}, with $c\s{co}$, resulting in 
\begin{equation}
\mu_3  = 2 \alpha c\s{co} \left( \ln \left(\frac{2 \ell }{r\s{co}} \right) + \frac{1}{4} \right) \, .
\label{eq_mu3}
\end{equation}

\subsection{counterion-wall interaction}

The fourth term is the attraction between the counterions, located at radial coordinate $r\s{ct}$, and the charged wall, which is given by 
\begin{equation}
e\s{ct-wall} = - 2 \lambda\s{B}  \sigma  \int_0^\ell \int_{-\pi}^\pi {r\s{p}}  \left( r_\text{p}^2-2 r\s{p}r\s{ct}\cos{\vartheta}+r_\text{ct}^2+z^2 \right)^{-1/2} \;  \text{d} \vartheta  \text{d} z  
\label{e_ctwall_int}
\end{equation}
which can be integrated with \textit{z}, and for large $\ell$ then results in
\begin{equation}
e\s{ct-wall} = - 4 \lambda\s{B}  r\s{p} \sigma \left(  \pi \ln \left( \frac{2 \ell }{r\s{p} } \right) - \frac{1}{2} \int_0^\pi \ln \left( \rho_\text{ct}^2-2 \rho\s{ct} \cos{\vartheta}+1  \right) \text{d} \vartheta   \right)
\end{equation}
where $\rho\s{ct}=r\s{ct}/r\s{p}$. The integral over $\vartheta$ is zero. The energy $e\s{ct-wall}$ leads to a contribution to the chemical potential given by
\begin{equation}
\mu_4 = - 4 \pi  \lambda\s{B} r\s{p}  \sigma  \ln \left( \frac{2 \ell }{r\s{p} } \right)  \, .
\end{equation}
This contribution from counterion-wall attraction exactly cancels the contribution from coion-wall repulsion, Eq.~\eqref{eq_e_co_wall_4}. Thus, the Coulombic contribution to the chemical potential of ions in a charged nanopore only depends on the three types of interactions between coions and counterions, and not directly on their interactions with the charged surface of the pore.

\subsection{counterion-counterion interaction}

The final contribution is the repulsion between the counterions. They are homogeneously distributed over a cylindrical surface of radius $r\s{ct}$. From the viewpoint of a given counterion, the Coulombic energy, analogous to Eq.~\eqref{e_ctwall_int}, is 
\begin{equation}
e\s{ct-ct} = \frac{ \alpha c\s{ct} }{2 \pi r\s{ct}} \;   \int_0^\ell \int_{-\pi}^\pi {r\s{ct}} \left( 2r_\text{ct}^2 \left( 1-\cos{\vartheta} \right) +z^2 \right)^{-1/2} \,  \text{d} \vartheta  \text{d} z  \, .
\label{e_ct_int}
\end{equation}
Integration of Eq.~\eqref{e_ct_int} with \textit{z} leads to an expression that for large $\ell$ simplifies to
\begin{equation}
e\s{ct-ct} = \frac{ \alpha c\s{ct} }{\pi} \left( \pi \ln \left( \frac{2 \ell }{r\s{ct} } \right) - \tfrac{1}{2} \int_0^\pi \ln \left( 2 - 2 \cos{\vartheta}  \right) \text{d} \vartheta  \right) \, .
\label{F_con_final}
\end{equation}
In Eq.~\eqref{F_con_final}, the integral over $\vartheta$ is again zero. Thus, we obtain 
\begin{equation}
e\s{ct-ct} = \alpha c\s{ct} \ln \left( \tfrac{2\ell}{r\s{ct}} \right)
\label{e_ct_con}
\end{equation}
which we multiply by $c\s{ct} $ and differentiate with $c\s{ct}$, resulting in
\begin{equation}
\mu_{5}= 2 \alpha c\s{ct} \ln \left( \tfrac{2\ell}{r\s{ct}}\right)   \, .
\end{equation}

\begin{figure}
\centering
\includegraphics[width=0.8\textwidth]{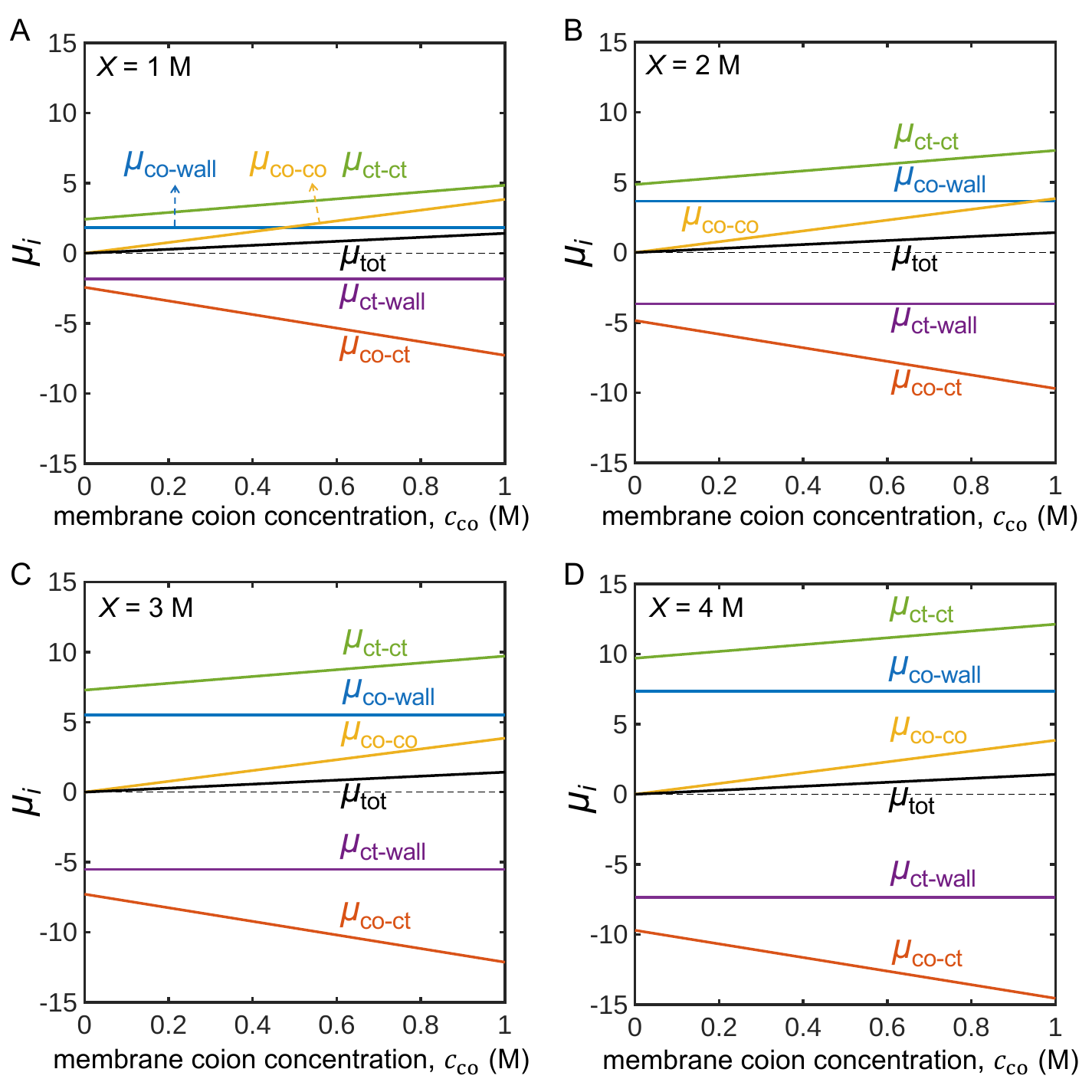}
\vspace{-0pt}
\caption{Contributions to the dimensionless chemical potential of an ion in a charged nanopore, for four values of the membrane charge density, \textit{X}, adding up to a total contribution, $\mu\s{tot}$. Calculations are based on $r\s{p} \! = \! 1$ nm and $a \! = \! 0.2$~nm. }
\label{fig_mu_terms}
\end{figure}

These five terms together fully define the Coulombic interactions of the ions with one another and with the pore wall, and we can add them together to a final expression that we discuss in the next section. In Fig.~\ref{fig_mu_terms} we show calculation results of these five contributions to the chemical potential, $\mu_1 \dots \mu_5$, as function of coion concentration in the membrane, as well as the total potential $\mu\s{tot}$. Fig.~\ref{fig_mu_terms} consists of four panels each with a different membrane charge density, and as can be observed, though individual terms, $\mu_i$, depend on \textit{X}, the total potential, $\mu\s{tot}$, is invariant with \textit{X}. The total potential, $\mu\s{tot}$, is also independent of the calculation length, $\ell$, as long as $\ell$ is high enough (see discussion in next section). However, the individual contributions do depend on $\ell$, increasing steadily in magnitude when $\ell$ increases. Thus, to construct a diagram such as Fig.~\ref{fig_mu_terms}, an arbitrary value of $\ell$ must be chosen. A value of for instance $\ell \! = \! 10$~nm or 100~nm, is then logical, but has the problem that all calculated lines are far apart, and the resulting $\mu\s{tot}$ cannot be noticed. Therefore, for the purpose of generating an insightful representation of the individual potentials, $\mu_i$, and their relation to $\mu\s{tot}$, we use in Fig.~\ref{fig_mu_terms} a much lower value, namely, $\ell=1$~nm.

\subsection{Contribution of Coulombic interactions to the mean chemical potential}

We add together $\mu_1$ to $\mu_5$, and obtain $\mu\s{tot}$. This $\mu\s{tot}$ we divide by 2 which results in an expression for the contribution of the Coulombic forces to the mean chemical potential, and this is given by
\begin{equation}
 \ln\gamma_{\text{m},\pm} = \tfrac{1}{2} \, \mu\s{tot} = \pi \lambda\s{B} \, r_\text{p}^2 \, c\s{co} \, N\s{av} \left( \ln \left( \frac{r\s{ct}}{r\s{co}}\right) + \tfrac{1}{4} \right) 
\end{equation}
or in shorter notation
\begin{equation}
\ln\gamma_{\text{m},\pm}  = \alpha \, c\s{co}  \left( \ln \left( \frac{r\s{ct}}{r\s{co}}\right) + \tfrac{1}{4} \right) \, .
\label{eq_lngamma_tot_2}
\end{equation}

This is our final result for the mean activity coefficient of a 1:1 salt in a charged nanopore. It is very interesting that this activity coefficient, $\ln\gamma_{\text{m},\pm}$, is a linear function of $c\s{co}$, and depends on ion radius, $a$, and pore radius, $r\s{p}$ (via a relationship such as for instance $r\s{ct} \! = \! r\s{p} \! -\! a$ and $r\s{co} \! = \! r\s{p} \! - \! 2a$), but is not directly dependent on membrane charge density, \textit{X}. Eq.~\eqref{eq_lngamma_tot_2} also shows that even though individual contributions to $\mu\s{tot}$ do depend on the calculation length, $\ell$, the resulting equation for $\mu\s{tot}$ does not. 

\newpage

We also assumed in the derivation that $\ell$ is large enough relative to the pore radius~$r\s{p}$. Analysing the integral over $\vartheta$ in Eq.~\eqref{eq_e_co_wall_2} for instance, we find that to have less than a 5\% error, we need $\ell$ to be at least 2.5 times larger than~$r\s{p}$. 

In the next sections we derive the extended Donnan equation, and implement the above-derived expression for the activity coefficient of ions in a charged nanopore.

\section{Analysis of extended Donnan model of the first and second kind }

The extended Donnan equation can be derived from consideration of the chemical potential of an ion in solution, and in the membrane. 
At chemical equilibrium, these two chemical potentials are the same. 
The dimensional chemical potential of an ion is $\overline{\mu}_{j,i}$ where index \textit{j} refers to the phase and \textit{i} to the ion. We consider the solution phase ($j\!=\!\infty$) and the membrane ($j\!=\!\text{m}$). 

In both phases we have an expression for $\overline{\mu}_{j,i}$ (unit J/mol) that in each case is given by a summation of various contributions such as
\begin{equation}
\overline{\mu}_{j,i} = \overline{\mu}_{\text{ref},i} + \overline{\mu}_{\text{id},j,i} + \overline{\mu}_{\text{el},j,i} + \overline{\mu}_{\text{aff},j,i} + \overline{\mu}_{\text{exc},j,i} + RT \ln{\gamma}_{j,i} \, .
\label{eq_mu_gen_1}
\end{equation}
The various terms in Eq.~\eqref{eq_mu_gen_1} are discussed in detail in ref.~\cite{Biesheuvel_2022} and are briefly summarized next. The reference term is mainly of importance for chemical reactions and cancels in a balance of the same ion distributing between two phases. Next are the ideal term, given by $\overline{\mu}_{\text{id},j,i}= RT \ln\left(c_{j,i}/c\s{ref}\right)$, and the electrostatic term, $\overline{\mu}_{\text{el},j,i}= z_i F V $, where \textit{T} is temperature (in K), \textit{R} the gas constant ($R \! = \! 8.3144$~J/mol/K), and the reference concentration is set to $c\s{ref} \! = \! 1$~mM. Next is an affinity-term related to all possible interactions of an ion with the medium in which it is dissolved. These interactions can be of a chemical or physical nature, and in general end up as a constant factor, equivalent to a solubility. 
The excess term describes volumetric interactions between the porous medium and the ions, and between ions amongst each other, i.e., it describes effects of occupancy and volume exclusion. The last contribution is the activity effect which relates to local Coulombic interactions between ions. Of course one can decide to include affinity and excess effects in this  activity correction, but this is generally not done, so we also treat them separately.

We insert the expressions for $\overline{\mu}_{\text{id},j,i}$ and $\overline{\mu}_{\text{el},j,i}$ in Eq.~\eqref{eq_mu_gen_1}. We furthermore include the definition of partitioning coefficients for affinity and excess  effects, $\Phi_{\text{aff},i}$ and $\Phi_{\text{exc},i}$, both of which are defined as 
\begin{equation}
\Phi_{k,i}= \exp \left( - \left( \overline{\mu}_{k,\text{m},i} - \overline{\mu}_{k,\infty,i}\right)/RT\right) \, .
\end{equation}
We also introduce the dimensionless electrical potential, $\phi = V / V\s{T}$ with the thermal voltage $V\s{T}=RT/F$, and we introduce the Donnan potential which is $\phi\s{D} = \phi_{\text{m}}-\phi_\infty$. We equate the chemical potential in solution with that in the membrane, i.e., $\overline{\mu}_{\infty,i}=\overline{\mu}_{\text{m},i}$, which leads to
\begin{equation}
\ln \left( c_{\infty,i} / c\s{ref} \right) + \ln \gamma_{\infty,i} = \ln \left( c_{\text{m},i} / c\s{ref} \right) +\ln \gamma_{\text{m},i} - \ln \left(  \Phi_{\text{aff},i} \right) - \ln  \left( \Phi_{\text{exc},i} \right) + z_i \phi\s{D} \, .
\label{eq_chem_pot_eq_1}
\end{equation}
We combine the partitioning coefficients for affinity and volume (excess) effects into a common partitioning coefficient for non-electrostatic effects, $\Phi_i$, and then Eq.~\eqref{eq_chem_pot_eq_1} is rewritten to
\begin{equation}
c_{\text{m},i} \, \gamma_{\text{m},i}= c_{\infty,i} \, \gamma_{\infty,i} \,  \Phi_{i} \, e^{-z_i\phi\s{D}}
\label{eq_ext_boltzmann}
\end{equation}
which is the extended Boltzmann equation including activity coefficients and the non-electrostatic partitioning coefficient, $\Phi_i$, which is a constant in the sense of not directly depending on ion concentrations. When $\Phi_i \! = \! 1$, the regular Boltzmann equation is arrived at~\cite{Schlogl_1966}. We can solve Eq.~\eqref{eq_ext_boltzmann} for multiple ions and combine with charge neutrality in solution and in the membrane, which is Eq.~\eqref{eq_EN}. Helpful in evaluating Eq.~\eqref{eq_ext_boltzmann} is that all ion concentrations are based on the open pore volume~\cite{Galama_2013,Kamcev_2015,Biesheuvel_Dykstra_2020}. It is not necessary to do that but then, if concentrations are defined per unit total membrane, a porosity correction is required in Eq.~\eqref{eq_ext_boltzmann}, with $\Phi=p$, where $p$ is porosity \cite{Biesheuvel_Dykstra_2020}.

Now we can go from the extended Boltzmann equation to the extended Donnan equation, which can be done for any mixture of salt ions of all kinds of valencies, but for more than two different ions, an analytical equation such as below is not easily found. 
In those cases it is better to solve the problem numerically. In this paper we only consider a single 1:1 salt, and then we do obtain analytical equations. 
We set up the extended Boltzmann equation for both ions, Eq.~\eqref{eq_ext_boltzmann}, and recombine to
\begin{equation}
c_{\text{m},+} \, c_{\text{m},-} \, \gamma_{\text{m},\pm}^2= c^2_{\infty} \, \gamma^2_{\infty,\pm} \, \Phi^2  
\label{eq_extD_3}
\end{equation}
where we introduce the mean partitioning coefficient $\Phi$, given by $\Phi=\sqrt{\Phi_{+}\Phi_{-}}$~\cite{Biesheuvel_2023}, and the mean activity coefficients, $\gamma_{\infty,\pm}$ and $\gamma_{\text{m},\pm}$, given by $\gamma_{j,\pm}=\sqrt{\gamma_{j,+} \gamma_{j,-}}$~\cite{Wright_2007,Biesheuvel_2020}. Thus the extended Donnan equation does not require the individual $\Phi_i$ or $\gamma_i$ to be the same for the two ions, but we only need to know the geometric average of each contribution. Note that the electrostatic term related to the Donnan potential, $\phi\s{D}$, has disappeared. The salt concentration, $c_\infty$, is the concentration both of anions and cations in solution. 
Eq.~\eqref{eq_extD_3} is the same as Eq.~(6) in ref.~\cite{Kamcev_2015}, and Eq.~(4) in ref.~\cite{Kamcev_2016}, except that in these papers, the authors set $\Phi$ to unity. Just as in refs.~\cite{Kamcev_2015} and \cite{Kamcev_2016} , we use $\Gamma$ as shorthand for the combined group $\gamma_{\infty,\pm}^2 / \gamma_{\text{m},\pm}^2$. When this group, $\Gamma$, does not depend much on salt concentration, $c_\infty$, we arrive at an extended Donnan equation of the first kind (assuming also that $\Phi$ is not dependent on salt concentration)~\cite{Galama_2013}. But when the $\Gamma$-function does have a direct or indirect dependence on salt concentration, we have an extended Donnan equation of the second kind~\cite{Kamcev_2016}.

\begin{figure}
\centering
\includegraphics[width=0.55\textwidth]{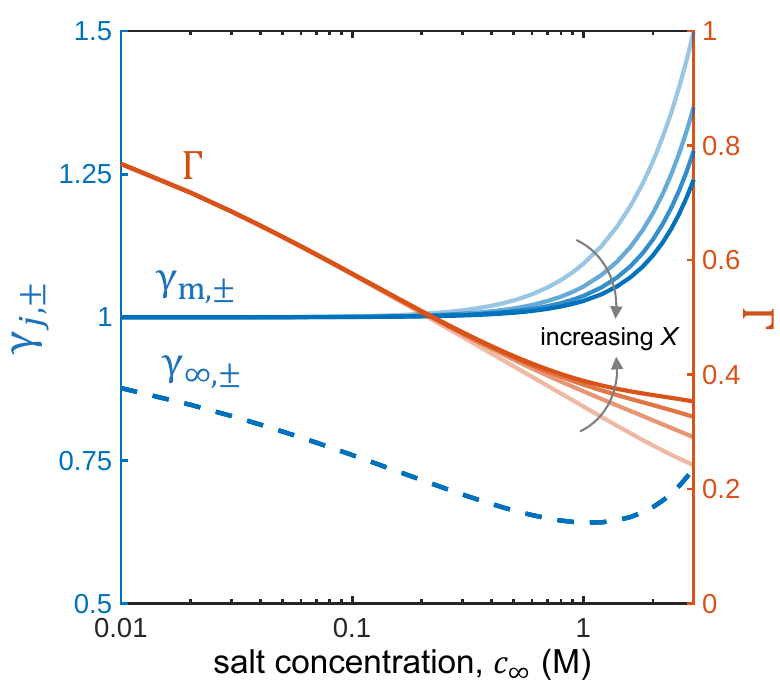}
\vspace{-0pt}
\caption{Analysis of mean activity coefficients of 1:1 salt solution in a membrane (in charged nanopores), $\gamma_{\text{m},\pm}$, and in outside solution, $\gamma_{\infty,\pm}$, in an extended Donnan calculation, as function of outside salt concentration, $c_\infty$. For membrane charge density, we use $X=1\text{,~}2\text{,~}3\text{,~or~}4$~M. The membrane activity coefficient, $\gamma_{\text{m},\pm}$, is constant and not dependent on membrane charge up to around $c_\infty = 0.5$~M, but beyond that, $\gamma_{\text{m},\pm}$ increases with \textit{decreasing} membrane charge density, because of an \textit{increasing} coion concentration in the membrane. As a consequence, the function $\Gamma=\gamma_{\infty,\pm}^2 / \gamma_{\text{m},\pm}^2$ steadily decreases with $c_\infty$ with a moderate dependence on \textit{X} beyond $c_\infty=1$~M.}
\label{fig_gamma_vs_cs}
\end{figure}

\newpage

We can also combine Eq.~\eqref{eq_extD_3} with charge neutrality, Eq.~\eqref{eq_EN}, and obtain  for the coion concentration in the membrane
\begin{equation}
c\s{co} = \sqrt{ \tfrac{1}{4} \, X^2 + \Gamma \, \Phi^2 \, c_{\infty}^2  } - \tfrac{1}{2} \, \left|X\right|  
\label{ext-Donnan}
\end{equation}
with the counterion concentration given by Eq.~\eqref{eq_EN}, and the Donnan potential calculated from~\cite{Biesheuvel_Dykstra_2020}
\begin{equation}
X = {2 \Phi c_\infty} \sinh \left(\phi\s{D} \right)
\end{equation}
which for $\Phi \! = \! 1$ simplifies to the regular Donnan equation~\cite{Schlogl_1966}.

Finally, we must set up expressions for the mean activity coefficients in solution. We use the extended Bjerrum equation, which is a simple equation that for 1:1 salts fits data very accurately up to salt concentrations beyond 1.5~M~\cite{Biesheuvel_2020}. The extended Bjerrum equation is
\begin{equation}
\ln \gamma_{\infty,\pm} =  - b  c_\infty^{1/3} -\tfrac{1}{4}  b^2 \cdot c_\infty^{2/3}  + 6  b^3 \cdot q  c_\infty
\label{eq_Bjerrum1}
\end{equation}
where \textit{b} is a constant factor that at room temperature is $b \!= \! 0.0605\text{~mM}^{-1/3}$. Furthermore, $q$ is a dimensionless size factor which for NaCl is $q \! = \! 0.19$ (for KCl it is $q \! = \! 0.125$). Instead of Eq.~\eqref{eq_Bjerrum1}, more complicated expressions such as the Pitzer model can also be used to achieve the same aim of correlating data for $\ln\gamma_{\infty,\pm}$ to $c_\infty$, but these models have a more elaborate structure.

We first analyse the complete extended Donnan model of the second kind, i.e., `Donnan-2', described by Eqs.~\eqref{eq_EN},~\eqref{eq_extD_3},~\eqref{eq_lngamma_tot_2}, and~\eqref{eq_Bjerrum1} in Fig.~\ref{fig_gamma_vs_cs} for parameter settings $r\s{p} \! = \! 1$~nm, $a \! = \! 0.2$~nm, and $\lambda\s{B}\!=\! 0.7$~nm as function of salt concentration, $c_\infty$, for different values of membrane charge density, \textit{X}, and we use the equations for $r\s{ct}$ and $r\s{co}$ suggested below Eq.~\eqref{eq_lngamma_tot_2}. The activity coefficient in solution, $\gamma_{\infty,\pm}$ follows the extended Bjerrum equation (dashed line) and the activity coefficient in the nanopores, $\gamma_{\text{m},\pm}$, follows Eq.~\eqref{eq_lngamma_tot_2}. We need to go beyond 0.5 M salt concentration to have an influence of \textit{X} on $\gamma_{\text{m},\pm}$, with $\gamma_{\text{m},\pm}$ increasing when \textit{X} decreases, which is because with decreasing \textit{X}, the coion concentration in the membrane increases. Because $\gamma_{\text{m},\pm}$ in all cases is higher than $\gamma_{\infty,\pm}$, the group $\Gamma=\gamma_{\infty,\pm}^2/\gamma_{\text{m},\pm}^2$ is always smaller than unity, and we notice how it gradually decreases with salt concentration. There is only a moderate effect of \textit{X} on $\Gamma$ only beyond $c_\infty = 0.5$~M, with $\Gamma$ increasing when \textit{X} increases.

\begin{figure}
\centering
\includegraphics[width=0.85\textwidth]{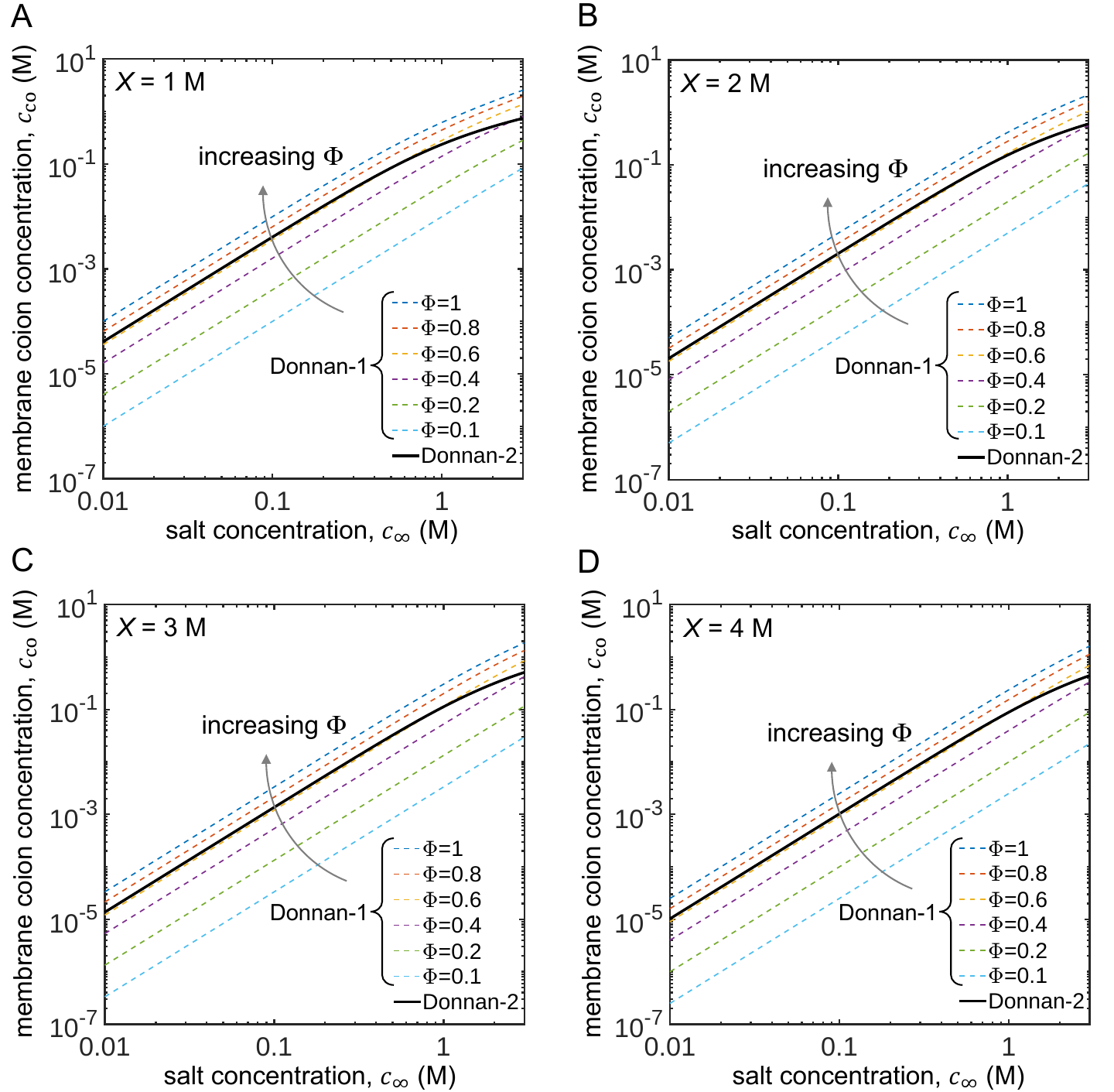}
\vspace{-0pt}
\caption{Comparison of extended Donnan model of the first kind (`Donnan-1'), using fixed $\Phi$ and $\Gamma \! = \! 1$ (dashed lines), with extended Donnan model of second kind (`Donnan-2'), based on Eqs.~\eqref{eq_lngamma_tot_2} and~\eqref{ext-Donnan} (solid lines). In this figure we leave out the activity effect in solution, thus $\Gamma=1/\gamma^2_{\text{m},\pm}$.}
\label{fig_cco_vs_cs_IDEAL}
\end{figure}

In Fig~\ref{fig_cco_vs_cs_IDEAL} we compare `Donnan-1' and `Donnan-2' in more detail by analysing calculation results for the predicted coion concentration in the membrane, $c\s{co}$, as function of salt concentration, $c_\infty$, and for different membrane charge densities, $X$. The dashed lines represent the extended Donnan model of the first kind (`Donnan-1'), for a range of values of $\Phi$, while $\Gamma=1$. The solid line is the Donnan model of the second kind (`Donnan-2') which includes the contribution to $\ln\gamma\s{m,\pm}$ according to Eq.~\eqref{eq_lngamma_tot_2}, but we leave out the activity effect in solution, i.e., we set $\ln\gamma\s{\infty,\pm}$ to unity. For $\Phi$ in the pore we include as an example a volume exclusion effect that is derived for a perfect sphere in a cylinder, 
given by $\Phi=\left(1-\lambda\right)^2$, with $\lambda \! = \! a/r\s{p}$~\cite{Biesheuvel_Dykstra_2020}. In all calculations, pore radius is $r\s{p} \! = \! 1$~nm and ion radius is $a \! = \! 0.2$~nm, and we use the equations for $r\s{ct}$ and $r\s{co}$ suggested below Eq.~\eqref{eq_lngamma_tot_2}. 

What Fig.~\ref{fig_cco_vs_cs_IDEAL} clearly shows is that for all membrane charge densities considered, there is a close match between Donnan-1 and Donnan-2 for the dependence of membrane coion concentration on salt concentration, $c_\infty$, up to $c_\infty \sim  0.8$~M, and beyond this range, Donnan-2 predicts a lower coion concentration than Donnan-1. But up to that point, the prediction of Donnan-2 is equal to that of Donnan-1. (In both models $\Phi=0.64$ is then implemented which relates to the sphere-in-cylinder formula for volume exclusion, $\Phi=\left(1-\lambda\right)^2$, that we use.) Notably, the Coulombic effect of the ion structure in the pore does not influence the model outcome, {at least not} up to a salt concentration of $c_\infty = 0.8$~M. 
For low enough salt concentration, according to Eq.~\eqref{ext-Donnan}, the `quadratic law' is valid, which describes that the membrane coion concentration depends on external salt concentration to the power 2 (this also implies that in a `log-log'~plot of $c\s{co}$ vs. $c_\infty$, the slope is 2)~\cite{Galama_2013}. This quadratic law is followed by Donnan-1 and as well by the present version of Donnan-2, because up to $c_\infty=0.8$~M, we have $\Gamma \! \sim \! 1$.

\begin{figure}
\centering
\includegraphics[width=0.85\textwidth]{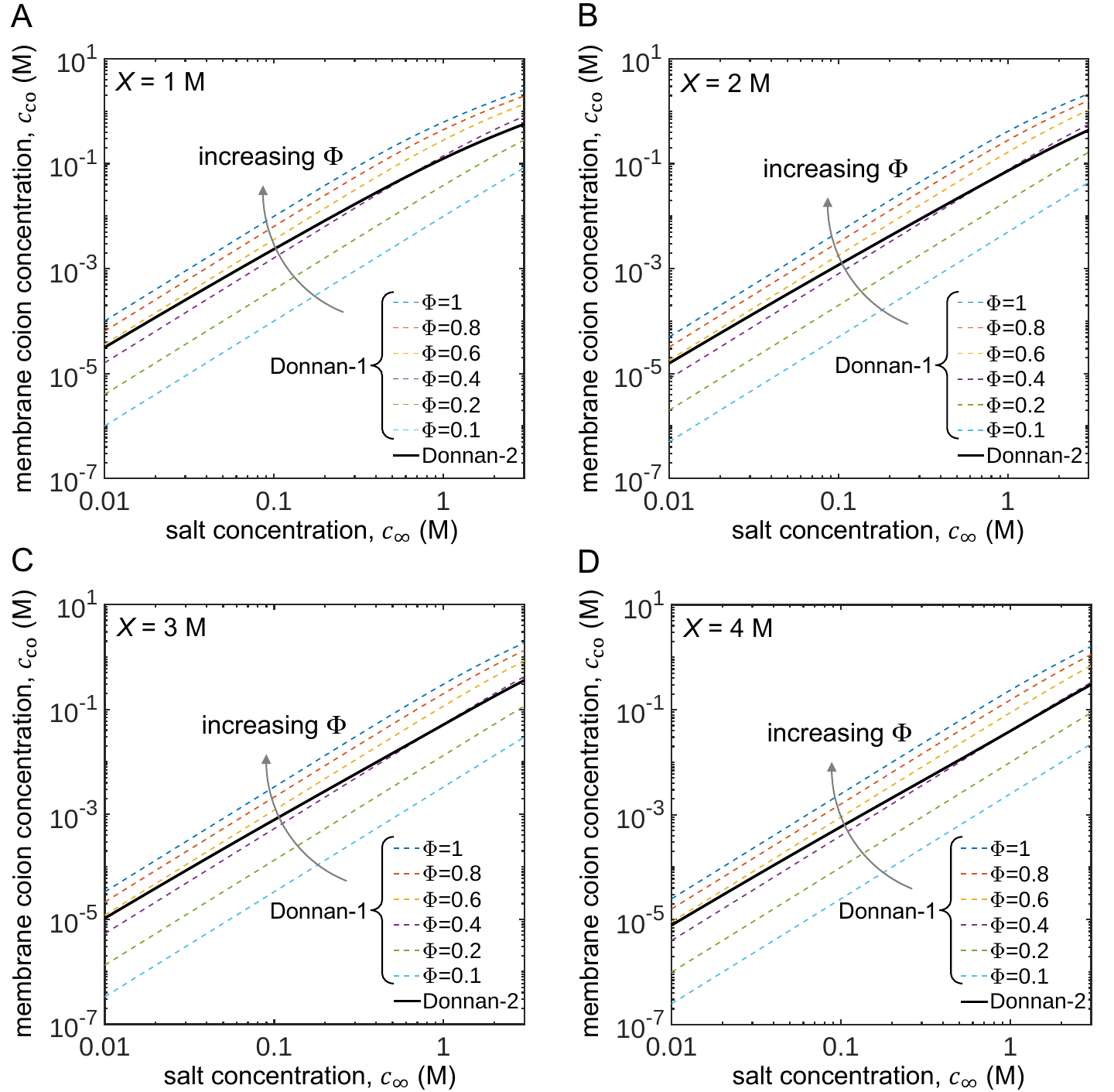}
\vspace{-0pt}
\caption{Comparison of extended Donnan model of the first kind (`Donnan-1'), using fixed $\Phi$ and $\Gamma \! = \! 1$; dashed lines), with extended Donnan model of second kind (`Donnan-2'), based on Eqs.~\eqref{eq_lngamma_tot_2},~\eqref{ext-Donnan}, and~\eqref{eq_Bjerrum1} (solid lines). Included is the activity effect in solution, thus $\Gamma=\gamma^2_{\infty,\pm} / \gamma^2_{\text{m},\pm}$.}
\label{fig_cco_vs_cs_NONIDEAL}
\end{figure}

When we do include the activity effect of the external solution, as described by Eq.~\eqref{eq_Bjerrum1}, we obtain the results presented in Fig.~\ref{fig_cco_vs_cs_NONIDEAL}. In this case, Donnan-2 theory has a slightly different slope at all salt concentrations than in Fig.~\ref{fig_cco_vs_cs_IDEAL}. However, the effect is not very prominent, with the apparent $\Phi$ slowly decreasing from $\Phi \! \sim \! 0.6$ to $\sim \! 0.4$ when salt concentration increases from 0.01 M to 1.0~M. Other than that gradual shift, the new Donnan-2 model stays close to Donnan-1 up to higher concentrations than in Fig.~\ref{fig_cco_vs_cs_IDEAL}. We can include the activity effect of ions in solution quite easily in the Donnan balance by combining Eqs.~\eqref{ext-Donnan} and~\eqref{eq_Bjerrum1}, which leads  to
\begin{equation}
c\s{co} = \sqrt{ \tfrac{1}{4} \, X^2 + \exp \left( - 2 b \, c_\infty^{1/3} -\tfrac{1}{2} \, b^2 \, c_\infty^{2/3}  + 12 \, b^3 \, q  \, c_\infty \right)  \Phi^2 \, c_{\infty}^2  } - \tfrac{1}{2} \, \left|X\right|   
\label{eq_ext_Donnan_2_simpl}
\end{equation}
which describes the solid lines in Fig.~\ref{fig_cco_vs_cs_NONIDEAL} up to $c_\infty = 1$~M closely. And this close fit is because we included $\ln\gamma_{\infty,\pm}$, and we do not need to include the activity model for ions in the membrane.

We can evaluate Eq.~\eqref{eq_ext_Donnan_2_simpl} in the low-$c_\infty$ limit, and we now find that the power law slope is no longer 2, as for the ideal Donnan model, but lower values are obtained, between 1.6 and 1.9, depending on \textit{X} and the function $\Phi=\left(1-\lambda\right)^2$.

\section{Conclusions}

In this paper we described a realistic model for the distribution of coions and counterions in a charged nanopore, and evaluated all Coulombic interactions of the ions with one another and with the pore wall. A concise final equation, Eq.~\eqref{eq_lngamma_tot_2}, was found for the Coulombic contribution to the mean chemical potential of ions in the pore that includes the dimensions of the pore and the coion concentration in the pore, but intriguingly predicts that membrane charge does not play an explicit role. This activity contribution to the chemical potential is small, and up to a salt concentration of $\sim0.8$~M does not influence the Donnan equilibrium. If we then neglect the activity effect of ions in solution, an extended Donnan model `of the first kind' is arrived at that includes a constant partitioning coefficient, but has no additional direct or indirect dependence on salt concentration. Such a constant factor can have many origins, including an affinity of ions with the membrane, and volume exclusion. 

When we include the Coulombic activity term in solution, we obtain an analytical solution for $c\s{co}$ in the membrane as function of external salt concentration, $c_\infty$, which is Eq.~\eqref{eq_ext_Donnan_2_simpl}. This equation predicts that in a `log-log' plot $c\s{co}$ depends linearly on $c\s{\infty}$. But the slope of this curve is less than the factor 2 that is arrived at in the extended Donnan model of the first kind, now varying between 1.6 and 1.9. Thus, our results show that Coulombic effects between ions in a nanopore do not matter much to the Donnan equilibrium between a pore and outside solution. Many other effects are much more important, such as the activity of ions in solution, and also chemical effects such as the dependence of membrane charge on local pH, taking just one example. These results are different from the condensation theory applied to charged nanopores~\cite{Kamcev_2015,Kamcev_2016,Kitto_2022}, in which Coulombic effects for ions in the pore were calculated to be attractive and strong, and consequently the related $\Gamma$-function was estimated to have large values, for instance $\Gamma \! = \! 5$ and beyond. Instead, we find Coulombic effects in the pore to be much smaller, and repulsive, and thus, according to our theory the $\Gamma$-function is less than unity.

\end{document}